\documentclass[twocolumn]{revtex4}
\usepackage{epsf}
\newcommand{\be}{\begin{equation}}
\newcommand{\ee}{\end{equation}}
\newcommand{\bea}{\begin{eqnarray}}
\newcommand{\eea}{\end{eqnarray}}

\begin{document}

\title{Jamming transitions and avalanches in the game of Dots-and-Boxes}
\author{Richard Metzler}
\affiliation{Institut f\"{u}r Theoretische Physik und
  Astrophysik, Universit\"{a}t W\"{u}rzburg, Am Hubland,
  D-97074 W\"{u}rzburg, Germany}
\author{Andreas Engel}
\affiliation{Institut f\"{u}r Theoretische Physik,
  Otto-von-Guericke-Universit\"{a}t Magdeburg, P.O. Box 4120,
  D-39016 Magdeburg, Germany}

\begin{abstract}
We study the game of Dots-and-Boxes from a statistical
point of view. The early game can be treated as a
case of Random Sequential Adsorption, with a jamming
transition that marks the beginning of the end-game.
We derive set of differential equations to make
predictions about the state of the lattice at the
transition, and thus about the distribution of
avalanches in the end-game. 
\end{abstract}

\maketitle
Real-life games have traditionally inspired research by 
mathematicians, economists, psychologists, computer scientists and also
physicists, starting with the development of probability
theory to handle games of chance 
\cite{Huygens:Ratiociniis,Bernoulli:Ars,Girenzer:Empire}.
Later fields of research were economic game theory
\cite{Jianhua:Games,Fudenberg:Learning}, 
where usually two or more players have to
make simultaneous and independent decisions without 
knowing what the other player is going to do, and 
combinatorial game theory
\cite{Berlekamp:Winning,Demaine:Playing}, where players
take alternating turns, and all information on the 
state of the game and the possible future moves is available.

The popular children's game of Dots-and-Boxes falls into the
latter category. The game is played on a rectangular lattice
(a checkered sheet of paper), and players take alternating
turns. At each turn, the active player occupies an edge. If
he thus occupies the fourth edge of at least one of the
two adjacent squares, the player continues the  turn by
placing another edge. Rules vary on whether a player 
must take a square if he has the chance -- the option not
to take it allows for a number of subtle moves (for a guide
to the end-game of Dots-and-Boxes, see
Ref. \cite{Berlekamp:Winning}, Vol. 2); however, for
simplicity, we demand that any chance to take a 
square must be used. The game ends when
all edges and squares are occupied. The player who took 
more squares wins.

The game can be separated into two distict phases: in 
the early game, players usually occupy edges more or 
less at random. However, they avoid placing the third edge
around any square, which would give the opponent the 
opportunity to score. Phase 1 ends when this is no longer 
possible: all free edges have at least one adjacent 
square with two occupied edges. This situation is
analogous to a jamming transition in models of random 
sequential adsorption (RSA). The main focus of this paper
is the modeling of the early game, using methods from 
RSA theory and Monte Carlo simulations, and to determine the
time and state of the game at the jamming transition.

In the end game, squares are rapidly filled: each edge
creates an opportunity to score, which often triggers another
opportunity, and another, until the avalanche is terminated
somehow.  The end-game 
is largely determined by nonlocal strategies (like figuring
out what the shortest possible avalanches are), and thus not
accessible to the methods used for the early game.  
However, the state of the game at the beginning of the
end-game limits the options available for the rest of 
the game, and allows for predictions of the minimum number of
turns until the end of the game, the average size of
avalanches, and so forth.

\section{The model}
\label{Sec-Model}
As a model for the game that can be both 
simulated with reasonable computational effort and 
handled analytically, we use the following
rules of behaviour for the players:
\begin{enumerate}
\item {\em Occupy squares if possible}: in accordance
  with the rules that prescribe a greedy strategy, the active player 
  searches the board for all free edges with an adjacent square 
  with exactly three occupied edges around it, picks one 
  of them at random, occupies it, and continues his turn. 
  This is repeated as often as possible.
\item {\em Create no opportunities for the opponent}: 
  if there are no
  squares to be taken, the player looks for free edges 
  whose adjacent squares both have less than two occupied
  edges. One of those is picked at random and occupied,
  and the turn ends.
\item {\em Minimize the opponent's score}:
  if both prescriptions (1) and (2) yield no suitable edge,
  the active player has to pick the third edge around some square, 
  thus giving his opponent a chance to take it according to 
  (1). He checks all edges to see how many points his
  opponent would gain from them. He picks one of 
  those that give away the smallest number of points, 
  occupies it, and ends the turn.
\end{enumerate}

To get rid of boundary effects, the lattice on which the
game is played is assumed to have periodic boundary conditions.
Furthermore, the analytical treatment is strictly valid
in the limit of infinite lattice size only.  

The lattice has $N$ squares. Accordingly,
the number of edges is $2N$. The number of turns is 
counted by $T$. The rescaled time $t= T/(2N)$ runs from
0 to at most 1; however, since more than one edge per turn is
occupied in the end-game, the game ends at times $t<1$.
 Two other useful quantities to describe
the system are the number $P$ of occupied edges (or
$p=P/(2N)$, the probability that a given edge is occupied),
and the number of filled squares $S$ (or $s=S/N$, respectively).

\section{The early game}
\label{Sec-Early}
As mentioned in the introduction, the early game can be 
treated as a special case of random sequential adsortion
(RSA) \cite{Evans:Random}. The basic idea of RSA is to 
deposit particles (atoms, dimers, or, in our case, edges)
on randomly chosen sites of a surface (often on a regular lattice)
unless this deposition violates restrictions posed by
particles that were adsorbed before. In Dots-and-Boxes,
the edges form a square lattice with a 
peculiar short-range three-particle repulsion. 

The usual procedure to treat RSA problems analytically is to
solve a set of coupled differential equations that describe the
probability of encountering the various possible
configurations of particles (for details, see
\cite{Evans:Random}). Unfortunately, this set of ODEs
is generally not closed, i.e., the equations for
small configurations include probabilities of larger 
configuration, which in turn depend on still larger
configurations. At some point, one has to neglect 
correlations and truncate the equations. Since the
number of terms increases dramatically with increasing
order of truncation, we will use the simplest approximation
that still captures the interaction correctly.

\begin{figure}
  \epsfxsize= 0.99 \columnwidth
  \epsffile{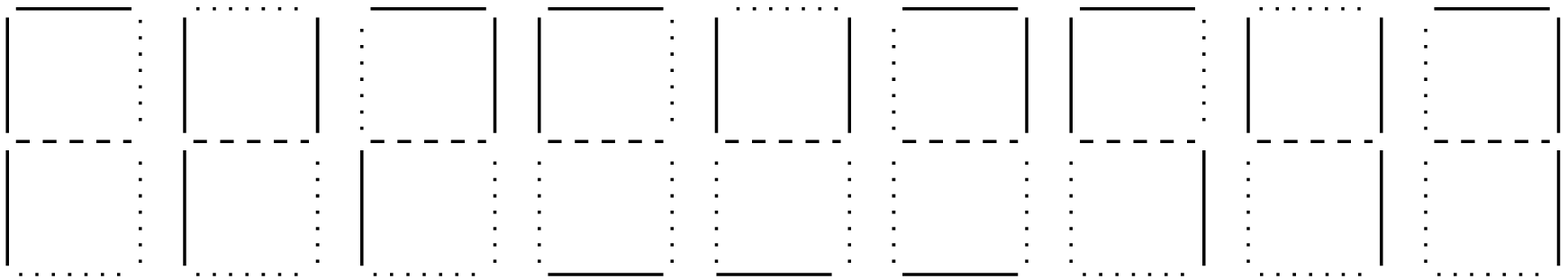}
  \caption{Free edges are classified according to the
number of occupied edges in adjacent squares. In all
cases shown here, the central edge (dashed line) has the indices 21.
All possible configuration with the same index $ij$ 
are assumed to have equal probabilities.}
\label{DG-configs}
\end{figure}

This approximation characterizes each free edge by two indices --
the first index $i$ for the number of occupied edges
surrounding the adjacent square above or to the left of the
considered edge, the second index $j$ for the occupation 
number of the square below or to the right (for example,
Fig.\ref{DG-configs} shows all configurations with index
$21$ around a horizontal edge). One can then 
count the number $F_{ij}$ of free edges with indices $ij$,
and determine their density $f_{ij} = F_{ij}/(2N)$.

All possible configurations with the same indices are
now assumed to have the same probability, and correlations
beyond nearest neighbours are neglected. For example, 
if we consider a free edge with indices $11$, we have no 
information on the free edges surrounding its adjacent
squares, except that one of their indices must be 1. 
Accordingly, we assume that the other index 
(let us say, $j$) follows 
a simple conditional probability, $\mbox{Prob}(j|i) = 
f_{ij}/\sum_k f_{ik}$. 

In the inititial phase of the game, there is an extensive
number of edges that can be occupied without giving the 
opponent an opportunity to score, following prescription 
(2) of the strategy. These are edges from the categories
00, 10, 01, and 11.
Now we can see what happens to the $f_{ij}$ if an
edge is occupied, i.e., a time step
$dt=1/(2N)$ elapses. The chosen edge (let us say, it
has indices $kl$) is occupied, and $f_{kl}$ is 
decreased by  $df_{kl}= 1/(2N)$. This means that
for all edges that can be chosen, the differential 
equation for $f_{kl}$ includes a loss term equal to 
the probability that an edge with index $kl$ is
picked: 
\be
\frac{df_{kl}}{dt} = -\frac{f_{kl}}{f_f} + \dots \mbox{ for
  } k,l \in \{0,1\},
\ee
where $f_f$ is the density of edges that can be 
occupied before the transition, 
\be
f_f = f_{00} + f_{01}+ f_{10} + f_{11}.
\ee
The free edges
in the squares next to the chosen edge must be 
updated: the index corresponding to the considered
square is increased by one. For instance, assume
that the upper/left adjacent square of the chosen edge
has occupation number 0 (which happens with probability
$(f_{00}+f_{01})/f_f$). Two of the three remaining edges
around that square will then have their second index increased
by one, whereas one will have its first index incremented.
This leads to gain and loss terms in the ODEs:
 \bea
f_{i0}&:&  -2 \frac{f_{i0}}{\sum_{k=0}^2 f_{k0}}
\frac{f_{00}+f_{01}}{f_f}; \nonumber \\
f_{0i}&:& - \frac{f_{0i}}{\sum_{k=0}^2 f_{0k}}
  \frac{f_{00}+f_{01}}{f_f} 
\nonumber  \\
f_{i1}&:&  +2 \frac{f_{i0}}{\sum_{k=0}^2 f_{k0}}
\frac{f_{00}+f_{01}}{f_f}; \nonumber \\
f_{1i}&:& + \frac{f_{0i}}{\sum_{k=0}^2 f_{0k}}
  \frac{f_{00}+f_{01}}{f_f}.
\eea
What follows is a rather tedious summation of 
terms for the possible combinations of indices.

The equations derived in this fashion can be 
simplified considerably by 
assuming the symmetry $f_{ij}= f_{ji}$, thus
keeping only the categories with $i\geq j$.
With the abbreviations
\bea
r_0 &=& (f_{00}+f_{10})/(f_{00}+f_{10}+f_{20}); \\
r_1 &=& (f_{10}+f_{11})/(f_{10}+f_{11}+f_{21}),
\eea
the resulting system of differential
equations looks as follows:
\bea
\frac{df_{00}}{dt} 
  &=& (-f_{00} -6 f_{00} r_0)/f_f;  \nonumber \\
\frac{df_{10}}{dt} 
  &=& (-f_{10} + 3 f_{00} r_0 - 3 f_{10} r_0 - 2 f_{10}
  r_1)/f_f ; \nonumber \\
\frac{df_{11}}{dt} 
  &=& (-f_{11} + 6 f_{10}r_0 - 4 f_{11} r_1)/f_f; \nonumber \\
\frac{df_{20}}{dt} 
  &=& (-3 f_{20}r_0 + 2 f_{10}r_1)/f_f;\nonumber \\
\frac{df_{21}}{dt} 
  &=& (- 2 f_{21} r_1 + 2 f_{11} r_1 + 3 f_{20}r_0)/f_f;\nonumber \\
\frac{df_{22}}{dt} 
  &=& 4 f_{21} r_1/f_f.
\eea
This can be solved numerically and compared to simulations.
The agreement is very good, but becomes slightly worse
close to the jamming transition -- the point where $f_f$
becomes zero and the early game ends. The accuracy could 
probably be improved by including probabilies of larger
configurations; however, it is not worth the effort, 
since the calculated numbers agree with results from real
play only in the order of magnitude anyway (see Section \ref{Sec-RealPlay}).
The predictions and numerical values for the
jamming time $t_J$ and the order parameters at that time
are given in Table \ref{DG-squaretab}:
\begin{table}[h]
\begin{center}
\begin{tabular}{c|c|c}
&\ \  theory\ \ \ \ \ & simulation \\ \hline
$t_{J}  $    & 0.4615 & 0.4657 \\
$f_{22} $ & 0.3244 & 0.3409 \\
$f_{21} $ & 0.0901 & 0.0846 \\
$f_{20} $ & 0.0169 & 0.0121
\end{tabular}
\end{center}
\caption{Order parameters at the jamming transition for
the square lattice.}
\label{DG-squaretab}
\end{table}

\begin{figure}
  \epsfxsize= 0.99 \columnwidth
  \epsffile{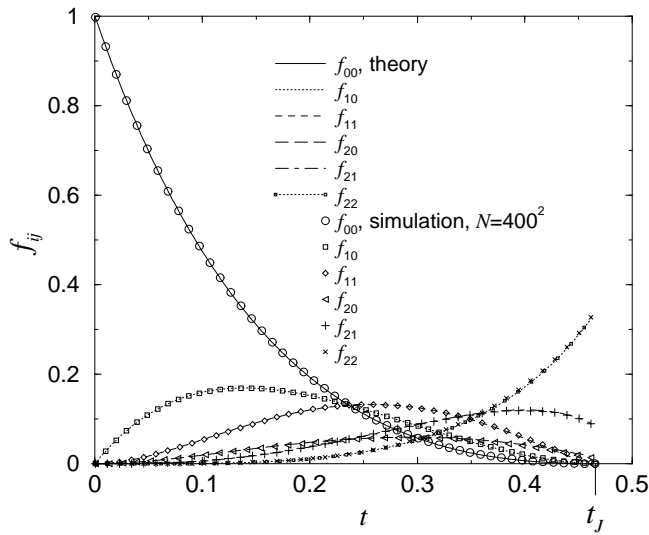}
  \caption{Comparison between the numerical solution of
the ODEs and a simulation with $N=400\times400$.}

\end{figure}
\section{The end-game}
When the last edge from the categories 00, 01, 10, and 11
has been taken, prescription (2) is no longer an option.
The game now alternates between prescriptions (1) and 
(3): Each player's turn begins by filling the squares of
the avalanche that his opponent has offered him by 
placing the third edge around some square (prescription
(1)). When all possible squares have been taken, the active 
player now determines the avalanche that his opponent must
take (prescription (3)). Since the length of the avalanche
triggered by placing an edge is not a local property of
that edge (and highly correlated to that of neighboring
edges), a description by differential equations 
analogous to the early game makes little sense. However, 
since the state of the system at the transition largely
determines the options of the players later on, we can make 
quantitative predictions about the end-game from the 
knowledge gained in Sec. \ref{Sec-Early}.

\begin{figure}
  \epsfxsize= 0.95 \columnwidth
  \epsffile{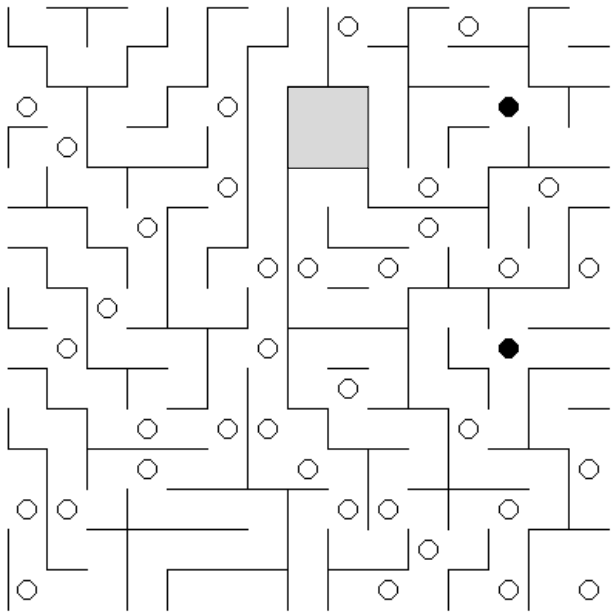}
  \caption{A typical state of the system at the jamming
    transition. Single defects are marked with empty 
    circles, double defects with full circles, and
    shading denotes a closed area.}
 \label{jamfig}
\end{figure}

Fig. \ref{jamfig} shows the
state of a game with $15\times15$ squares at the jamming transition.
The squares are segments of a tunnel if they have two
occupied egdes. They represent tunnel branchings
(or {\em single defects}, marked by empty circles) 
if they have only one edge, and tunnel crossings 
(or {\em double defects}, marked by full circles) 
if they have none. 

Thus, three edges from the $f_{21}$-category 
form a single defect, whereas four $f_{20}$-edges
make up a double defect. The density of defects can be 
calculated directly from the order parameters at
the transition.

As an additional check, one can make sure that 
the total number of edges (both occupied and 
free) add up to $2N$, and that the number of
squares (tunnels and defects) add up to $N$. 
This leads to the following equations:
\bea
t_J + 2 f_{20} + 2 f_{21} + f_{22} &=& 1; \\
3 f_{20} + (10/3) f_{21} + 2 f_{22} &=& 1 ,
\eea 
which are fulfilled for both the analytical and
the experimental values.

This enables us to make some statements about the avalanches
or chains of occupied squares that occur in the late game.
An avalanche started in a tunnel fills the tunnel and 
ends at the defect edges on both sides of the tunnel.
Thus, with $4 N (f_{21} + f_{20})$ defect edges, 
there are 
at least $N_A = 2 N (f_{21} + f_{20})$ different 
potential avalanches at the time of the transition.
(For now, we neglect avalanches in closed areas and other 
complications.) We can also calculate the number of
tunnel segments (squares with occupation 2)  
from the order parameters:  
$N_T= N - (4/3)N f_{21} - N f_{20}$. 
This yields an average length of the tunnel segments 
of $N_T/N_A \approx 4.5$ for the values 
of $f_{ij}$ from the simulation, as given above.
 
Note that in an analogy to the ``waiting time
paradoxon'', the average avalanche length becomes
larger than the mentioned value of 4.5 if avalanches 
are started at randomly chosen edges rather than randomly chosen
tunnel segments, because longer tunnels include more 
edges and are thus chosen with higher probability:

Let us assume that the probability distribution 
$P_{av}(l)$ of tunnel lengths 
$l$ follows an exponential with a decay constant
$1/l^{\ast}$. Since $l\geq 1$, the normalization 
constant is $ \exp(1/l^{\ast})-1$, and the 
average value is
\be 
\langle l \rangle_{av} = \sum_{l=1}^{\infty} l (e^{1/l^{\ast}}-1)
e^{-l/l^{\ast}} =  \frac{1}{1- e^{-1/l^{\ast}}}. 
\ee
With the mentioned value of $\langle l \rangle_{av}=4.5$,
one gets $l^{\ast}\approx 4.0$ and 
$P_{av}(l) \approx 0. 284 \exp(-l/4.00)$.  
Since each tunnel avalanche of $l$ squares length has 
$l+1$ edges where it can be started, the probability
distribution of avalanche lengths averaged over free
edges follows the form $P_{ed}(l)\propto (l+1)
\exp(-l/l^{\ast})$.
\begin{figure}
  \epsfxsize= 0.99 \columnwidth
  \epsffile{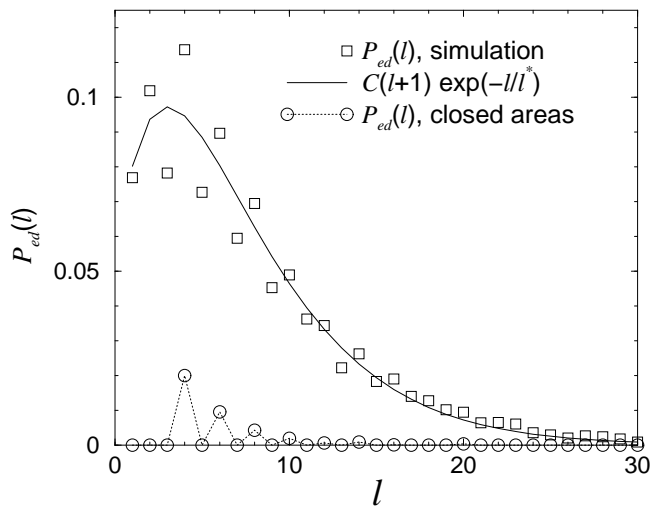}
  
  \caption{Probablility distribution $P_{ed}(l)$ of avalanche lengths
    $l$ at the jamming transition, averaged over randomly chosen free
    edges. Simulations average over 40 runs with $N=50\times50$. }
\label{DG-P_l}
\end{figure}

This agrees fairly well with simulations, as seen in 
Fig. \ref{DG-P_l}. However, there is a preference for
even avalanche lengths, which can partly be explained
with the presence of {\em closed areas}. These are areas 
that contain no defects and are separated from the rest of
the board by occupied edges. They include an even number
$\geq 4$ of squares. The probability of an edge being in
a closed area of size $l$ is shown in Fig. \ref{DG-P_l}
(open circles). Even if it is taken into account, even 
avalanches are more likely, for reasons that are still unclear. 

The density of defects at the transition allows for a 
prediction of the total number of turns in the end-game:
When an avalanche is terminated at a defect, that defect is
turned into a tunnel segment (if it was a single defect
before) or into a single defect (if it was a double defect).
Therefore the number of defect-terminated avalanches $N_{DTA}$ is 
half the number of single defects plus the number of
double defects: 
\be
N_{DTA} = ((2/3)f_{21} + f_{20})/N \approx 0.069N.  
\ee
This means that the time difference from $t_J$ to the end of
the game $t_E$ should be at least
\be
t_E-t_J \geq N_{DTA}/(2N) = 0.034.
\ee
However, this is really only a lower bound on the time 
found in simulations, $t_E-t_J \approx 0.054$. 
The reason for the deviation is the existence of avalanches that do 
not change the number of defects, namely,  
avalanches in closed areas. These areas are not necessarily
present at the beginning of the end-game. Instead, they may 
initially be half-closed areas: areas that are not quite closed,
but connected to the rest of the system by a single defect. 
Depending on whether the avalanche is 
started in the tunnel outside the half-closed area or inside
it, it is either turned into a closed area, or it is 
filled, the defect is removed, and the adjacent tunnel is
filled as well. Since it is usually desirable to give the
opponent as few points as possible, most of the half-closed
areas in real play will be turned into closed areas, and
then filled.

\begin{figure}
  \epsfxsize= 0.99 \columnwidth
  \epsffile{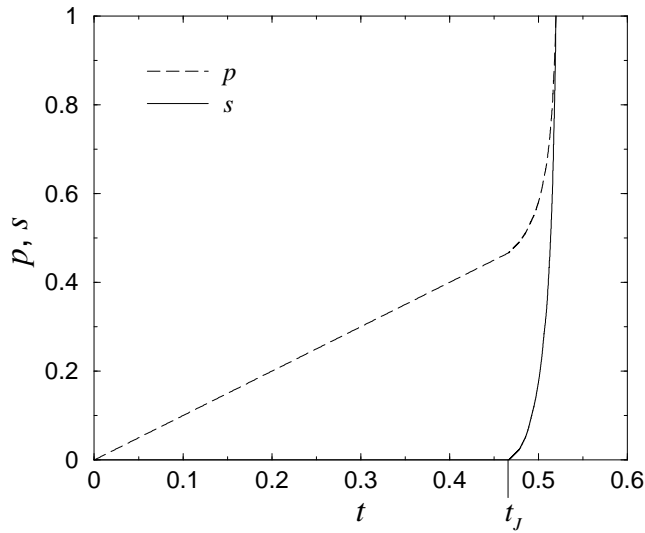}
  \caption{Time development of fraction of occupied
edges ($p$) and squares ($s$) in a simulation on 
an $80\times 80$ lattics.}
\label{DG-endgame}
\end{figure}

Apart from these exceptions, avalanches tend to get
dramatically longer as the end-game goes on 
(see Fig. \ref{DG-endgame}). This
is due to two effects: first, small avalanches are triggered
earlier than larger ones due to prescription (3),
and thus removed; second, avalanches that stop at a single
defect turn it into a tunnel segment, merging two 
potential avalanches into one. 

\section{Other lattices}
The game can be played on lattices other than 
the square, as long as there is a notion of 
an edge separating two cells, and placing 
a single edge is not enough to make scoring possible.
The simplest case is the triangular lattice,
where there are only single defects, and the
relevant order parameters in the early game are 
$f_{00}$, $f_{10}$, and $f_{11}$. The corresponding
differential equations are 
\bea
\frac{df_{00}}{dt} &=& -1 - 4\; \frac{f_{00}}{f_{00}+f_{10}};\\
\frac{df_{10}}{dt} &=& 2\;\frac{f_{00}-f_{10}}{f_{00}+
  f_{10}};\\
\frac{df_{11}}{dt} &=& 4\; \frac{f_{10}}{f_{00} + f_{10}}.
\eea  
They can even be solved analytically by introducing 
a rescaled time $\tau$ with $d\tau =
 (f_{00}+f_{10})^{-1}dt$,
and solving the remaining system of linear ODEs in $\tau$  with
constant coefficients. One gets
  \bea
f_{00}(\tau) &=& e^{-4\tau} (2 - e^{\tau}); \label{DG-triangsol1} \\
f_{10}(\tau) &=& 2 e^{-4\tau} (e^{\tau}-1);
\label{DG-triangsol2} \\
f_{11}(\tau) &=& 2 e^{-4\tau} - (8/3) e^{-3\tau} + 2/3.
\eea  
Using Eqs. (\ref{DG-triangsol1}) and (\ref{DG-triangsol2}),
$t$ can be calculated: 
\be
t(\tau) = (1-e^{-3\tau})/3; \ \tau(t) = -(1/3) \ln(1-3t).
\label{DG-triangsol4}
\ee
\begin{figure}
  \epsfxsize= 0.99 \columnwidth
  \epsffile{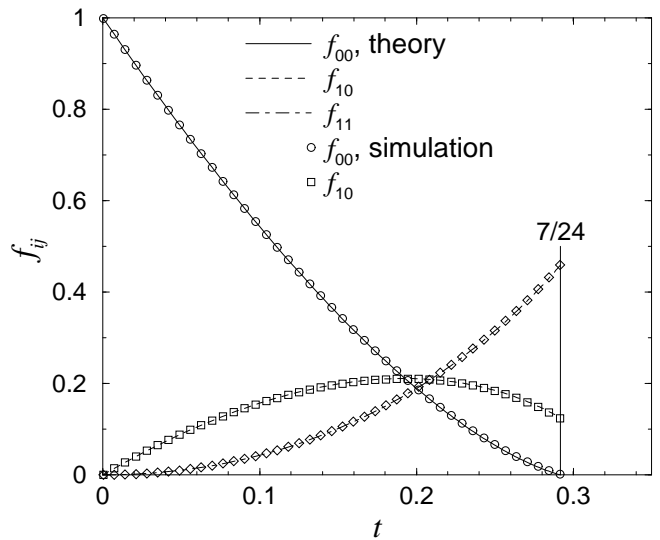}
  \caption{Comparison between the analytical solution 
(Eqs. (\ref{DG-triangsol1})--(\ref{DG-triangsol4})) of
the ODEs and a simulation on a triangular lattice with $N=20000$.}
\label{DG-triang_plot}
\end{figure}
Again, agreement between theory and simulation is very
good, as seen in Fig. \ref{DG-triang_plot}.
The predicted and observed values for the jamming transition and the
order parameters are given in Table \ref{DG-3angtab}. 
\begin{table}[h]
\begin{tabular}{c|c|c}
&theory & simulation \\
\hline
$t_J$ & 7/24= 0.291667 & $0.2930\pm 0.0002$ \\
$f_{10}$ & 1/8 =0.125 & $0.1209 \pm 0.0003$ \\
$f_{11}$ & 11/24 = 0.45833 & $0.4655 \pm 0.0002$ 
\end{tabular}
\caption{Order parameters at the transition for the triangular lattice}
\label{DG-3angtab}
\end{table}

Other possible lattices include hexagonal and 
three-dimensional cubic lattices. In the latter case,
edges correspond to faces of unit cubes. Interestingly,
the differential equations are the same 
for both the hexagonal and the 3D-cubic lattice, since
both have six edges/faces surrounding each hexagon/cube. 
Although structurally simple, the equations involve fifteen 
order parameters and are not written out for the sake of brevity.

Of course, the whole range from single to quadruple 
defects can occur; however, multiple defects are
rarer than single ones, as seen in Table \ref{DG-3dtab}.
\begin{table}[h]
\begin{tabular}{c|c|c|c}
&theory & simulation 3D & simulation hexagonal\\
\hline
$t_J$ &  0.6367 & $0.6381 \pm 0.0001$ &    $0.6351 \pm 0.0001$\\
$f_{40}$ &\  0.00017\ &\  $0.00010 \pm 0.00003\ $ &\
$0.00030 \pm 0.00003 $\ \\
$f_{41}$ & 0.0027  & $0.0022\pm 0.00001$ & $0.0036 \pm 0.0001$\\ 
$f_{42}$ & 0.0177  & $0.0162 \pm 0.0002$ & $0.0198 \pm 0.0002$\\
$f_{43}$ & 0.0577  & $0.05723  \pm 0.0002$ & $0.0579 \pm 0.0002$\\ 
$f_{44}$ & 0.2064 & $0.2105 \pm 0.0002$ & $0.2018 \pm 0.0002$
\end{tabular}
\caption{Order parameters for the 3D cubic and hexagonal lattice.}
\label{DG-3dtab}
\end{table}

Results from simulations confirm the picture predicted by 
calculations, with the usual deviations on the order of $10^{-3}$.

Is the game still interesting on other lattices?
Disregarding the practical difficulties of playing
on a 3D-cubic lattice, all basic mechanisms of the
game still work, including closed and half-closed
areas. A rough estimate shows that the
initial avalanche length (averaged over possible 
avalanches) is 4.8 for the triangular lattice and
3.6 for the 3D-cubic and hexagonal lattice,
similar to the square lattice. 
We therefore expect that real-life games on other
lattices would not be much different from 
Dots-and-Boxes on regular square lattices. 

\section{Comparison to real play}
\label{Sec-RealPlay}
We let some coworkers play a computer
version of Dots-and-Boxes (with periodic boundary 
conditions and $N=10\times 10$) to see if their style of
play is well described by the assumptions in Section
\ref{Sec-Model}. Generally speaking, human players did not
place edges at random in the early game; instead they
tended to add edges to existing structures. In some
cases, this led to a significantly lower number of
defects, and thus longer avalanches. One pair
of players chose to get rid of the periodic 
boundary conditions by drawing a frame around the
board early in the game.

\begin{figure}
  \epsfxsize= 0.99 \columnwidth
  \epsffile{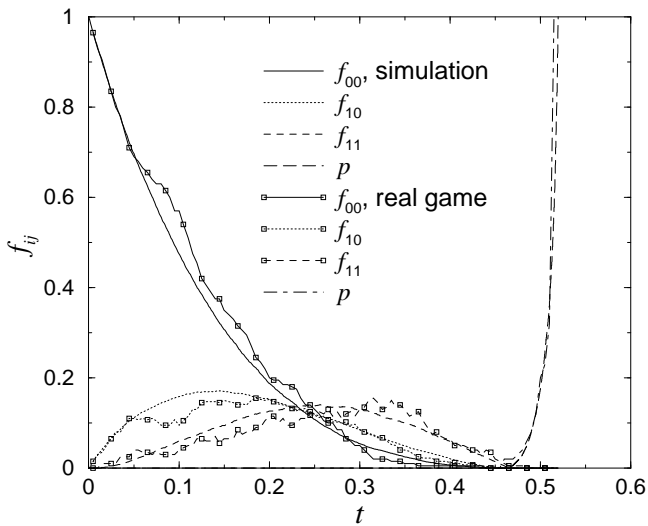}
  \caption{Comparison between a simulation with
    $N=80\times80$ and a real-life game on a
    $10\times10$-lattice. 
    To avoid confusion, only the curves for $f_{00}$, $f_{10}$,
    $f_{11}$, and the density of filled squares $p$ is shown
    in this figure -- see also Fig. \ref{DG-AvsP2}.}
\label{DG-AvsP1}
\end{figure}

\begin{figure}
  \epsfxsize= 0.99 \columnwidth
  \epsffile{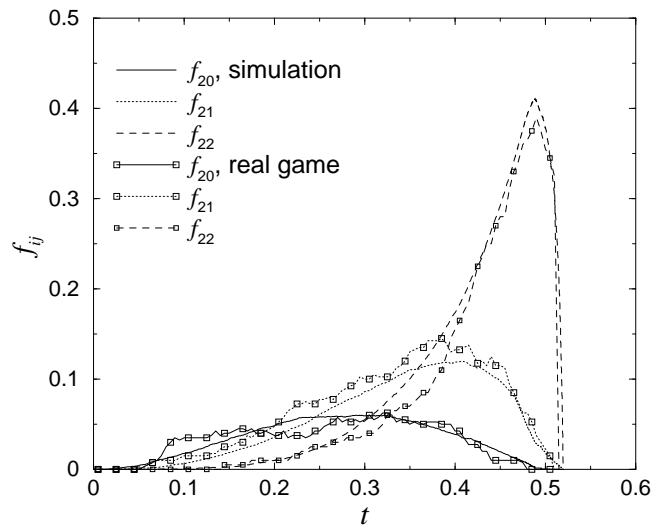}
  \caption{The remaining order parameters of the game shown 
    in Fig. \ref{DG-AvsP1}.}
\label{DG-AvsP2}
\end{figure}

Nevertheless, some games showed quantitative similarities to 
our theoretical predictions. The order parameters from
one of these games is shown in Figs \ref{DG-AvsP1} and
\ref{DG-AvsP2}.

Our test players  did not try tactical 
subtleties like adjusting the number of avalanches in order to get the
last (and presumably longest) one. They were usually 
happy if they avoided blatant mistakes.

Of course, one could include human tendencies and 
extend the model to include {\em cooperative 
sequential adsorption} \cite{Evans:Random}, where
edges are preferrably placed next to edges occupied before.
However, since this could not describe all human players
with the same set of parameters and would probably give
no qualitative new insights, the usefulness of this
extension is questionable. 
\ \\
\section{Summary and conclusion}
We gave a statistical treatment of the game of
Dots-and-Boxes, using some simplifying assumptions
for the behaviour of the players. In the early game, 
since a finite fraction of edges can be chosen, a
mean-field description given by a system of coupled
differential equations works well. It makes predictions
about the point where  avalanches start and
the degree of geometrical frustration at that point. 
The same scheme works for all kinds of regular lattices;
the relevant quantity is the number of edges or faces 
around a cell, such that hexagonal and 3D-cubic lattices
are described the same equations. 

These predictions allow for statements regarding the
statistics of avalanches, as well as the total number
of turns in the end-game. The presence of closed
and half-closed areas makes the situation more complicated;
unfortunately, they cannot be captured by the approximations
made in the calculation of the early game.

While results from calculations give good agreement with 
simulations, human players have various habits that 
cannot be easily included in an all-encompassing
mean-field treatment (``I like making corners. They 
look nice.'') Thus, while our analysis has yielded
some insight in the underlying processes of Dots-and-Boxes,
quantitative agreement with human play is not always 
satisfactory.

\begin{acknowledgments}
R. M. acknowledges finacial support by the GIF. We also like to
thank our test players, and Wolfgang Kinzel and Michael
Biehl for pointing us to relevant literature.
\end{acknowledgments}

\end{document}